\documentclass[useAMS,usenatbib]{mn2e}
\usepackage{longtable}
\usepackage{epsfig}
\usepackage{graphicx}
\usepackage{amssymb}

\title{The evolution of the luminosity-temperature-mass relations of hot gas in Chandra clusters at 0.4 $< z <$ 1.4.}

\author[Iu. Babyk, I. Vavilova]{Iu. Babyk$^{1,2,3}$\thanks{E-mail:
babikyura@gmail.com}, I. Vavilova$^{1}$\thanks{E-mail: iriv@mao.kiev.ua}\\
$^{1}$ Main Astronomical Observatory NAS of Ukraine, Zabolotnogo str., 27, 03650, Kyiv, Ukraine\\
$^{2}$ Dublin Institute for Advanced Studies, 31 Fitzwilliam Place, Dublin 2, Ireland \\
$^{3}$ Dublin City University, Dublin 9, Ireland \\
}

\begin{document}

\date{Accepted \underline{\hspace*{2cm}}. Received \underline{\hspace*{2cm}}; in original form \underline{\hspace*{2cm}}}

\pagerange{\pageref{firstpage}--\pageref{lastpage}} \pubyear{0000}

\maketitle

\label{firstpage}

\begin{abstract}
We analyzed the luminosity-temperature-mass of gas ($L_{X} - T - M_{g}$) relation for sample of galaxy clusters that have been observed by the $Chandra$ satellite. We used 21 high-redshift clusters ($0.4 < z < 1.4$). 

We assumed a power-law relation between the X-ray luminosity of galaxy clusters and its temperature and redshift $ L_{X}~\sim~(1~+z)^{A_{L_{X}T}}~T^{\beta_{L_{X}T}}$. We obtained that for an $\Omega_{m} = 0.27$ and $\Lambda = 0.73$ universe, $A_{L_{X}T} = 1.50 \pm 0.23$, $\beta_{L_{X}T} = 2.55 \pm 0.07$ (for 68~\% confidence level). 

Then, we found the evolution of $M_{g} - T$ relation is small. We assumed a power-law relation in the form $M_{g}~\sim~(1~+~z)^{A_{M_{g}T}}~T^{\beta_{M_{g}T}}$ also, and we obtained $A_{M_{g}T} = -0.58 \pm 0.13$ and $\beta_{M_{g}T} = 1.77 \pm 0.16$. We also obtained the evolution in $M_{g} - L_{X}$ relation, we can conclude that such relation has strong evolution for our cosmological parameters. We used $M_{g}~\sim~(1~+~z)^{A_{M_{g}L_{X}}}~L^{\beta_{M_{g}L_{X}}}$ equation for assuming this relation and we found $A_{M_{g}L_{X}} \approx -1.86 \pm 0.34$ and $\beta_{M_{g}L_{X}} = 0.73 \pm 0.15$ for $\Omega_{m} = 0.27$ and $\Lambda = 0.73$ universe.

In overal, the clusters on big redshifts have much stronger evolution between correlations of luminosity, temperature and mass, then such correlations for clusters at small redshifts. We can conclude that such strong evolution in $L_{X} - T - M_{g}$ correlations indicate that in the past the clusters have bigger temperature and higher luminosity.

\end{abstract}


\section{Introduction}\label{sec:1}

The correlations between different types of X-ray properties of galaxy clusters are useful method to obtain information about global properties of clusters. The observation of the diffuse, X-ray emitting medium (ICM) of galaxy clusters provides quantities like its mass ($M_{g}$), temperature ($T$), and X-ray luminosity ($L_{X}$). The analysis of the scaling relation between these physical properties gives considerable insight into the physical processes in the ICM. There are difficulties for theoretical predictions  of the evolution of these scaling relations (e.g. \citealt{Norman:10}). 

First of all, if concerns with some previous researches correlations between X-ray luminosity and temperature (\citealt{David:93}, \citealt{Markevitch:98}, \citealt{Mushotzky:97}). It is known that there is a correlation between the mass of hot intragalactic gas, temperature and X-ray luminosity in clusters. Such a correlation have the small scatter, that indicates on the similar formation history  for all clusters.

However, more detail researches show that the observational correlation value is at odds with theoretical predictions. The most famous example is the slope $ L_{X} \sim T^{2.7} $ for hot clusters \citealt{Markevitch:98}, while the theory provides $ L_{X} \sim T^{2} $ \citealt{David:93}. Such differences can be explained that heated intergalactic medium play an important role non gravitational processes as early heating of massive supernovae \citet{Cavaliere:97} or radiation cooling and the associated active star formation \citet{Voit:01}. The evolution of the correlation ratios cause ought to be important information that helps to choice between similar models. In addition, the correlation value at high redshifts is useful for cosmological studies based on the $L_{X}-T-M_{g}$ evolution of clusters, as they can connect the observed X-ray luminosity in terms of characteristics of clusters, such as temperature or mass \citealt{Borgani:01}.

We notice some works on the $L_{X}-T$ ratio for galaxy clusters at the large redshifts. By Mushotzky with so-authors in series of papers ({\citealt{Mushotzky:97a}, \citealt{Mushotzky:96}, \citealt{Mushotzky:97}, \citealt{Mushotzky:00}) analyzed this ratio for large sample of distant clusters observed ``telescope'' ASKA and not found the evolution in $L_{X}-T$ ratio. However, later observations  (\citealt{Arnaud:02}, \citealt{Novicki:02}) were existence of such an evolution as $ L_{X} (z) \sim (1 + z)^{A} $, then $ A = $ 1.3-1.5. \citet{Borgani:01} measurign parameters of 7 clusters at $ z> $ 0.5, concluded about some possible weak evolution, $ A <$ 1. A similar conclusion, that made by \citet{Holden:02}, measurements for 12 clusters of galaxies. Thus, the situation remained be unclear.

Value $ L_{X} - T $ for clusters has considerable static variation. Dispersion in $L_{X}-T$ for close clusters can be significantly reduced, if in determining the luminosity and temperature of clusters to exclude the central region (\citealt{Markevitch:98},(\citealt{Finoguenov:01}, \citealt{Finoguenov:07}).

We notice that results from different observational methods of mass measurements are not consistent with one another and with the simulation results (e.g., \citealt{Horner:99}, \citealt{Finoguenov:01}, hereafter FRB). In general, X-ray mass estimates are about 80 \% lower than the predictions of hydro-simulations. 

Another recent observational finding is the possible existence of a break in the $ T - M_{g}$ relation. By use of resolved temperature profile of X-ray clusters observed by ASCA, FRB have investigated $T - M_{g}$ relation in the low-mass end and find that $M \propto T^{\sim 2}$, compared to $M \propto T^{\sim 3/2}$ at the high mass end. 

In our sample is about 20 clusters at $ z > $ 0.4 with a large exposure time, which allows to determine the temperature of clusters with an accuracy better 10-15 \%. Using these $Chandra$ data we were able to accurately determine the evolution of the correlation relations between luminosity, temperature and mass of intracluster gas accumulations at $ z > $ 0.4. 

We derived constraints on the mass-temperature relation of gas in galaxy clusters from their observed luminosity-temperature relation. We parameterized the $M_{g} - T$ relation as a single power-law, and derive the best-fit values of their amplitude and slope from the observed $L_{X} -T$ relation. Also, we derived $L_{X}-M_{g}$ relation for our sample, we used power-low to describe this relation. 

In Section 2 we show the main steps of the data analysis and describe the sample. 

In Section 3 we give the methods to obtain the total mass and mass of hot gas. In Section 4, 5 and 6 we show the results of our measurements, discussion, and conclusion of our work, respectively.

In this paper we used the cosmological parameters: $H_{0} = 73~  km~  s^{-1} Mpc^{-1}$ and matter density $\Omega_{m}=0.27$. Errors are given at the 68\% (1$\sigma$) confidence level.

\section{Sample and Data processing}

The main observational parameters of the sample galaxy clusters are presented in Tab. \ref{tab1}. The table (colums 1-8) consists of the name of clusters, the redshift, the number of observation, the exposition time (before cleaning), the name of detectors, and the values of column density, which were described by \citet{Dickey:90}, and coordinates (RA, DEC) (which were taken from NED\footnote{http://nedwww.ipac.caltech.edu}). The Chandra observations were obtained using the Advanced CCD Imaging Spectrometer (ACIS). The list of events (level 1) was produced by the Chandra pipeline processing, and then was reprocessed using the CIAO software package. VFAINT mode was used to improve the rejection of cosmic ray events. 

We used the Chandra CIAO software package, version 4.2, for data reduction, including the latest maps and calibration products CalDB    4.5.1 . The main data processing steps were provided using techniques discussed by \citet{Babyk:12a}a, \citet{Babyk:12b}b, \citet{Babyk:12c}c, \citet{Babyk:12d}d, \citet{Babyk5:12}e in our previous research. 

In brief, we extracted annular spectra centered on the X-ray peak with different widths to ensure similar statistics in the background-subtracted spectra. The X-ray peaks were determined with the X-ray images from which all point sources were removed. The average of the outermost radius is about 1 Mpc. We have generated ARF and RMF files using the $mkarf$ and $mkrmf$ command in CIAO tool. Approximation of the spectra was carried out in the range 0.5 - 7.0 keV. 

Temperatures clusters determined by the spectrum collected within a radius of $R_{200}$, this area always contains at least 75\% of full radiation accumulation. In the approximation of the spectra was believed that the interstellar absorption in the galaxy corresponds to the thickness of neutral hydrogen derived from radio observations. The metallicity of intergalactic plasma was considered equal to 0.3 solar value, unless the quality of data did not allow to determine it directly from the observed spectrum. 

Normalization of the theoretical spectrum, which produces the best approximation, can give us the conversion factor between the number of photons from the source and X-ray luminosity for energy range and extent of issue $EM = \int n_{e} n_{p} dV $. This ratio allows to determine the emission of intergalactic gas. All luminosity is measured within the radius $R_{200}$. The central region with a strong peak brightness in the center, probably influenced by radiative cooling. In such cases, the central regions (100 kpc) were excluded, as the measurement of luminance and spectral analysis (\citealt{Markevitch:00}).

\begin{table*}
\centering
\fontsize{7}{7}
\selectfont
\caption{\fontsize{9}{9} \selectfont Summary of the Chandra observations (colums 1-8) and result of galaxy clusters (colums 9-11). Colums list the target name, redshift, number of ID observation, net exposure time, detector used, column density and coordinates (taken from NED), also the list of result of galaxy clusters: temperature, luminosities and masses, respectively.}
\bigskip
\label{tab1}
\begin{tabular}{llccccccccc}
\hline
  &   &   &  &  & $n_{H}$, &  &  &  & L, & $M_{g}$ \\
Cluster&$z$ &ObsID & $t_{exp}$, &Detector &10$^{20}$ & RA&DEC &T &10$^{45}$ & 10$^{13}$\\ 
&&&ks&& cm$^{-2}$&(J2000)&(J2000)& keV &  erg s$^{-1}$&  $M_{\odot}$\\
\hline
MACSJ0159.8-0849 & 0.405&3265&18.00&ACIS-I&2.08&01 59 48.0 &-08 49 00& 9.11$_{-0.83}^{+0.81}$ &8.46$_{-0.88}^{+0.84}$&21.33$_{-3.11}^{+3.12}$ \\
MACSJ0647.7+7015&0.59&3196&18.85&ACIS-I&5.63&06 47 45.9 &+70 15 03& 9.97$_{-1.22}^{+0.97}$ &10.11$_{-1.03}^{+1.04}$&57.25$_{-5.33}^{+4.89}$\\
MACSJ0744.8+3927&0.697&536&19.69&ACIS-I&5.68&07 44 51.8 &+39 27 33& 8.11$_{-0.61}^{+0.59}$ &3.01$_{-0.17}^{+0.33}$&22.94$_{-5.38}^{+6.72}$\\
MACSJ1311.0-0311 &0.49&3258&15.00&ACIS-I&1.88&13 11 00.0 &-03 11 00 & 8.55$_{-0.37}^{+0.29}$ &7.36$_{-0.48}^{+0.76}$&40.41$_{-10.72}^{+8.64}$\\
MACSJ1423.8+2404 &0.545&4195&113.40&ACIS-I&2.38&14 23 48.3 &+24 04 47& 7.04$_{-0.53}^{+0.51}$ &2.22$_{-0.19}^{+0.25}$&44.91$_{-6.41}^{+5.78}$\\
MACSJ2129.4-0741&0.589&3199&17.69&ACIS-I&4.84& 21 29 26.0 & -07 41 28& 8.21$_{-0.74}^{+0.71}$ &2.10$_{-0.10}^{+0.21}$&37.05$_{-2.16}^{+1.95}$\\
RCSJ1419.2+5326&0.64&3240&10.03&ACIS-S&1.18&14 19 12.0 &+53 26 00& 5.05$_{-0.36}^{+0.38}$ &0.64$_{-0.05}^{+0.06}$&15.40$_{-3.20}^{+3.21}$\\
RXJ0848.7+4456&0.570&927&126.74&ACIS-I&2.66& 08 48 47.2 &+44 56 17& 2.45$_{-2.30}^{+2.31}$&0.33$_{-0.04}^{+0.01}$&3.10$_{-0.57}^{+0.74}$\\
RXJ1347.5-1145&0.451& 2222&93.9 &ACIS-S &4.85 &13 47 32.0 & -11 45 42& 11.48$_{-0.29}^{+0.28}$ &18.33$_{-1.18}^{+1.28}$& 17.03$_{-3.05}^{+3.01}$\\
CLJ1226.9+3332&0.890&3180&32.12&ACIS-I&1.37&12 26 58.0 &+33 32 54& 6.44$_{-0.60}^{+0.58}$ & 3.02$_{-0.31}^{+0.29}$&39.45$_{-3.28}^{+3.27}$\\
ISCS J1438+3414 & 1.41 & 10461 & 150.00 & ACIS-S& 2.01& 14 33 22.4 & +22 18 35.4 & 4.99$_{-0.22}^{+0.24}$ &0.46$_{-0.05}^{+0.04}$& 2.44$_{-0.45}^{+0.35}$\\
RCSJ0224-0002&0.778&4987&90.15&ACIS-S&2.92&02 24 00.0 &-00 02 00& 5.27$_{-0.46}^{+0.55}$&1.02$_{-0.28}^{+0.21}$&9.46$_{-0.38}^{+0.75}$\\
RCSJ0439-2904&0.951&3577&77.17&ACIS-S&2.63&04 39 38.0 &-29 04 55& 5.55$_{-0.38}^{+0.75}$&1.19$_{-0.19}^{+0.17}$&8.58$_{-0.47}^{+0.57}$\\
RCSJ1107-0523&0.735&5825&50.12&ACIS-S&4.24&11 07 22.80& -05 23 49.0& 8.33$_{-0.84}^{+0.98}$&3.27$_{-0.38}^{+0.35}$&21.47$_{-3.18}^{+2.99}$\\
RCSJ1620+2929&0.87&3241&37.13&ACIS-S&2.72&16 20 09.40& +29 29 26.0& 2.18$_{-0.32}^{+0.29}$&0.22$_{-0.04}^{+0.02}$&13.18$_{-1.10}^{+1.39}$\\
RCSJ2318+0034&0.78&4938&51.12&ACIS-S&4.13&23 18 30.67 & +00 34 03.0& 8.11$_{-0.93}^{+0.88}$&4.16$_{-0.18}^{+0.42}$&18.36$_{-2.11}^{+3.09}$\\
RCSJ2319+0038&0.904&5750&21.18&ACIS-S&4.16 &23 19 53.00& +00 38 00.0& 7.55$_{-0.77}^{+0.78}$&2.18$_{-0.27}^{+0.23}$&17.33$_{-2.18}^{+1.98}$\\
RXJ0849+4452&1.26&945&128.45&ACIS-I&2.50&08 53 43.6 & +35 45 53.8& 5.44$_{-1.18}^{+1.77}$&0.97$_{-0.09}^{+1.01}$&3.32$_{-0.47}^{+0.56}$\\
RXJ0910+5422&1.106&2227&84.20&ACIS-I&2.35&09 10 45.36&+54 22 07.3& 6.74$_{-1.16}^{+0.95}$&1.55$_{-0.10}^{+0.17}$&2.37$_{-0.26}^{+0.33}$\\
RXJ1113.1-2615&0.730&915&105.95&ACIS-I&5.47&11 13 05.2 &-26 15 26 & 5.61$_{-0.44}^{+0.57}$&1.22$_{-0.21}^{+0.19}$&10.15$_{-2.18}^{+0.98}$\\
RXJ1221.4+4918 & 0.700& 1662& 80.13& ACIS-I& 1.45& 12 21 24.5 &+49 18 13& 5.88$_{-0.36}^{+0.66}$ &1.16$_{-0.17}^{+0.13}$& 12.11$_{-1.46}^{+0.99}$\\
\hline
\end{tabular}
\end{table*}

The spectra were analysed in Xspec (version 12.6) environment (\citealt{Arnaud:96}), using the MEKAL\footnote{MEKAL is the model which describes an emission from hot diffuse plasma (ICM)} code (\citealt{Kaastra:93}; incorporating the Fe-L calculations of (\citealt{Liedahl:95}) with WABS\footnote {The WABS is a parameter which describe the galactic absorption (\citealt{Dickey:90}).} parameter.

The annular spectra were modeled in order to determine the deprojected X-ray temperature and other parameters, under the assumption of spherical symmetry. We have used routine method DSDEPROJ (\citealt{Sanders:07}, \citealt{Russell:08}) to obtain  ``deprojected'' spectra. After deprojection, we determined the temperature of clusters and build the surface brightness profile for each cluster. 

For each annulus we determined the temperature, $kT$, and parameter $norm \sim \int n_{e}n_{H}dV$ which is proportional to the electron, $n_e$ and hydrogen, $n_H$, concentrations. Note, that the other parameters of model ($Z,z,$ etc) were fixed. Using these parameters, we have built the surface brightness profiles which fitted by $\beta$ model. The surface brightness profiles are measured in the 0.5-7 keV energy band, which provides an optimal ratio of the cluster and background flux in the Chandra data. 

 \begin{figure*}
 \begin{minipage}{0.45\linewidth}
 \epsfig{file=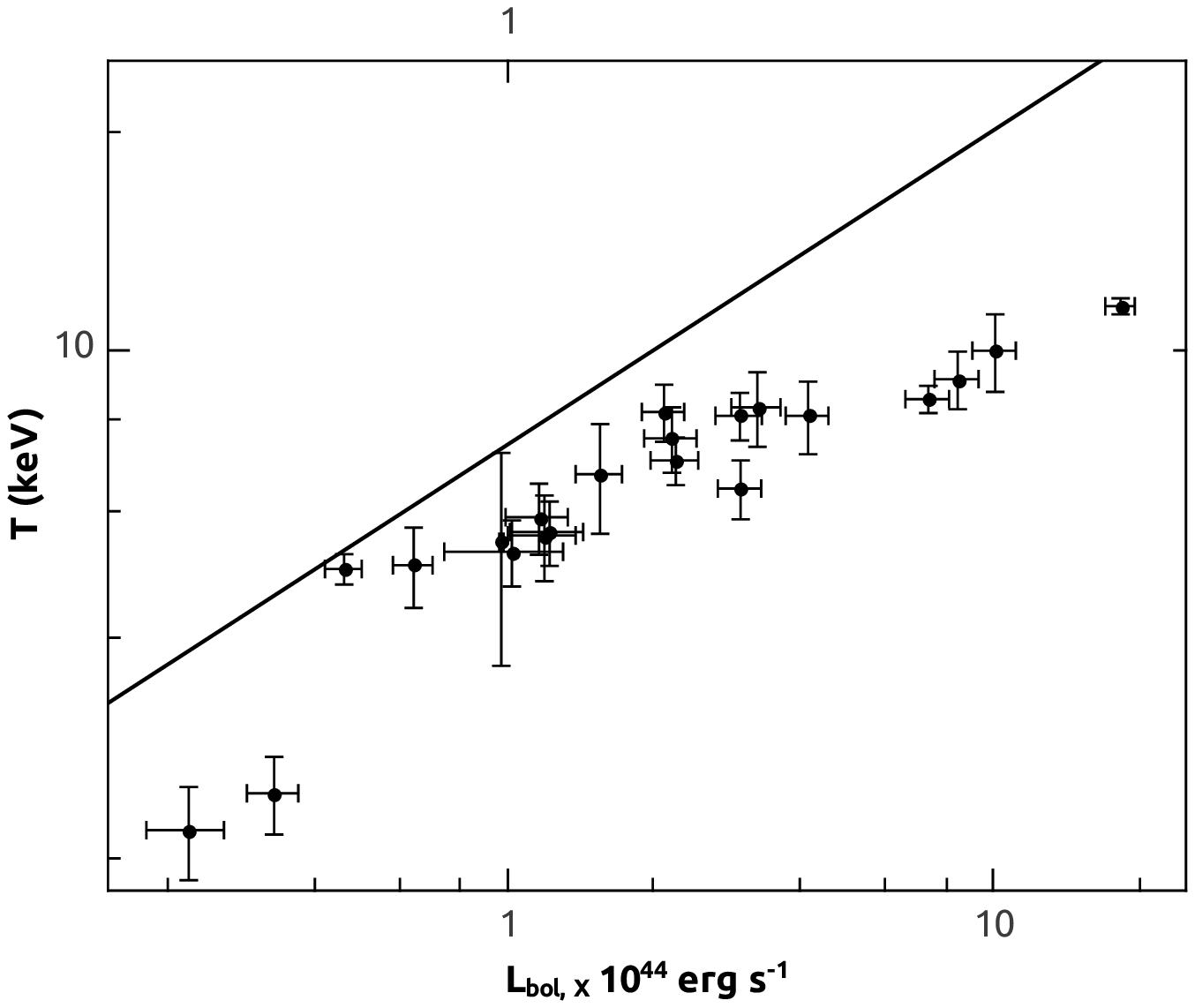, width=1.\linewidth}
 \end{minipage}
 \begin{minipage}{0.45\linewidth}
 \epsfig{file=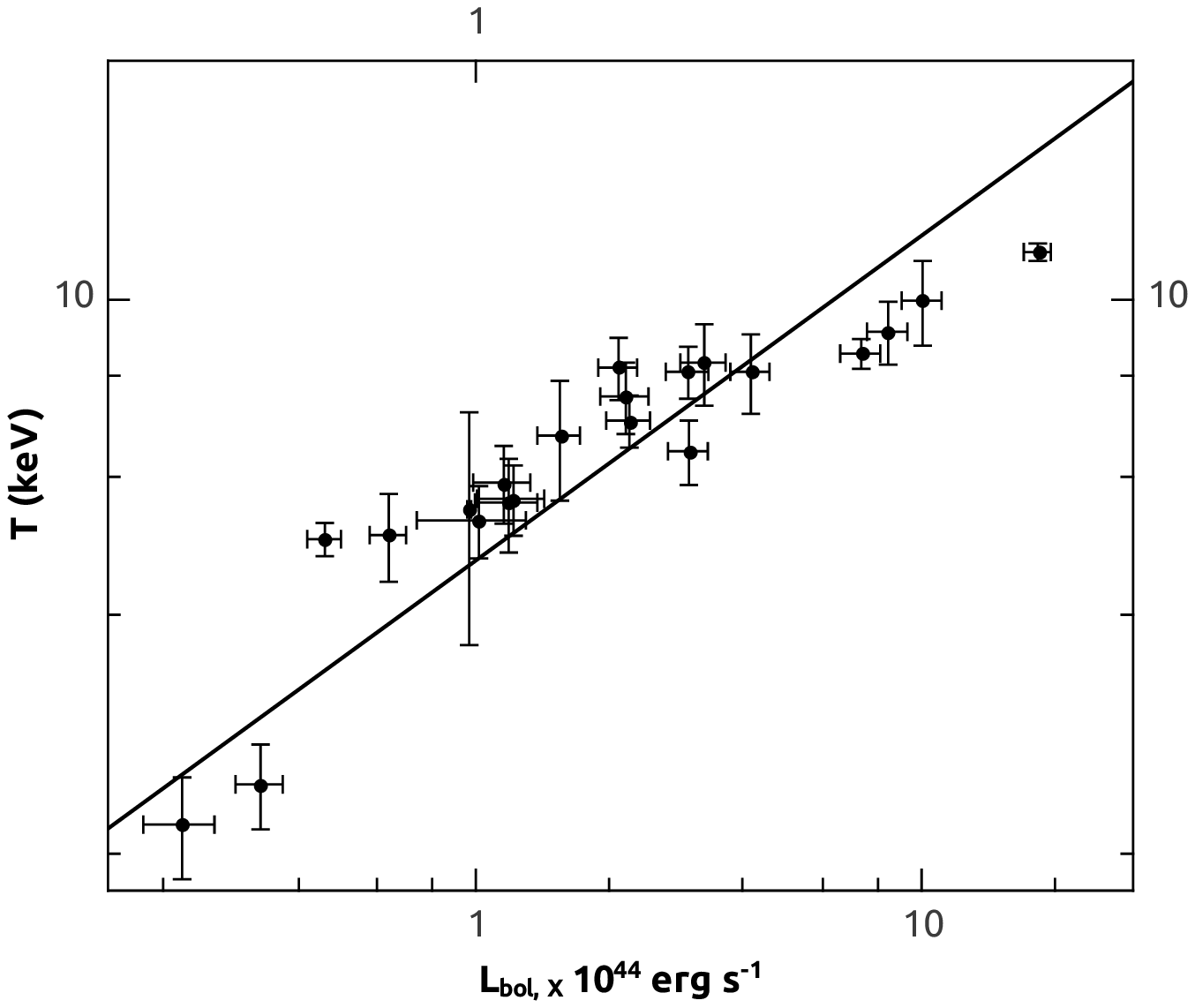, width=1.\linewidth}
 \end{minipage}
 \caption{The relation between temperature and luminosity of clusters. The solid line on left image represents $L - T$ relation for nearer clusters. On the right image the solid line moved on $(1 + z)^{-1.53}$.}\label{lt}
 \end{figure*}

\section{Methods}
\subsection{Galaxy cluster mass profile}
The gas in a cluster is trapped by the cluster's gravitational potential well. If the hot gas is supported by its own pressure against gravitational infall, it must obey the equation of hydrostatic equilibrium. The intra-cluster gas in most clusters appears to be in
approximate hydrostatic equilibrium,
\begin{equation}
\frac{dP}{dr} = -\frac{G M(r) \rho_{gas}(r)}{r^{2}},
\label{dpdr}
\end{equation}
where $P$ is the pressure of the gas, $G$ is the gravitation constant, $\rho$  is the density of the gas and $M$ is the total mass inside the sphere of radius $r$. We used the assumptions of spherical symmetry of clusters. Since the density and pressure of hot gas have very low values, we can use the ideal gas law, 
\begin{equation}
P = \frac{\rho_{g}kT_{g}(r)}{\mu m_{p}},
\label{p}
\end{equation}
where $\mu$ is the mean molecular weight of the gas, and $m_p$ the proton mass. Conbining equation (\ref{dpdr}) and equation (\ref{p}), the total mass of cluster as a function of the projected radius is,
\begin{equation}
M_{T}(r) = -\frac{kT(r)r}{G \mu m_{p}} \left( \frac{d \ln \rho_{g} }{d \ln r} + \frac{d \ln T}{d \ln r}\right).
\label{mtr}
\end{equation}

Assuming an isothermal, spherical gas cloud, with temperature $T$, equation (\ref{mtr}) becomes,
\begin{equation}
M_{T} = -\frac{k T r}{G \mu m_{p}} \left( \frac{d \ln \rho_{g} }{d \ln r} \right),
\end{equation}
where $r$ is the outer radius of the cluster and $\mu$ is constant with radius, since the chemical composition of the gas is expected to be uniform throughout the cluster. 

\subsubsection{Modeling of hot gas emission} 
Assuming an isothermal spherical gas cloud in hydrostatic equilibrium and assuming the volume density of galaxies follows a so called King profile (\citealt{King:72}), the density profile of X-ray emitting gas can be approximated by an isothermal $\beta$ model (\citealt{Cavaliere:76}),
\begin{equation}
\rho_{g}(r) = \rho_{0} \left( 1 + \left( \frac{r}{r_{c}}\right)^{2}\right)^{-3/2\beta},
\label{rog}
\end{equation}
where $\rho_{g}(r)$ is the density of the gas as a function of the cluster's projected radius $r$, $r_{c}$ is the core radius and $\rho_{0}$ is the density at the cluster's center. The values of $\beta$ and $r_{c}$ are obtained by analysis of the X-ray surface brightness profile for each cluster.

\subsubsection{X-ray surface brightness profile}

X-ray image analysis allows the determination of the X-ray surface brightness profile. Assumption that the hot gas is isothermal leads to a $\beta$ model for the cluster's brightness profile,
\begin{equation}
S(r) = S_{0} \left( 1 + \left( \frac{r}{r_{c}}\right)^{2}\right)^{-3\beta + 1/2} + C, 
\label{sr}
\end{equation}
where $S(r)$ is the X-ray brightness as a function of the projected cluster's radius $r$. $S_{0}$, $r_{c}$ , $\beta$ and $C$, are free parameters in fitting the model to the X-ray brightness profile. In general the X-ray surface brightness profile is well approximated by the $\beta$ model.

There are cases when the standard $\beta$ model leads to underestimating the brightness in the central region of the clusters. The central X-ray excess is one of the first pieces of evidence of cooling flows in galaxy clusters (\citealt{Jones:84}). For simplicity, in our study we use a single $\beta$ model (\citealt{Roncarelli:06}). In this paper, the X-ray surface brightness profile of each cluster in our sample and the fitted $\beta$ model were obtained from the original X-ray images.

\subsection{The gas mass}
The mass of the hot gas in the ICM is determined by the integration of the gas density profile (Eq.~\ref{rog}), over the volume of the cluster within a defined radius $R_{200}$,
\begin{equation}
M_{g} = \int_{200} \rho dV.
\label{mg}
\end{equation}
Equation (\ref{mg}) can be written as,
\begin{equation}
M_{g} = 4\pi \rho_{0} \int_{0}^{R_{200}} r^{2} \left( 1 + \left( \frac{r}{r_{c}}\right)^{2}\right)^{-3/2\beta} dr.
\label{mg1}
\end{equation}

The value of $\rho_{0}$ is derived from combining the results obtained from fitting a model to the X-ray spectrum, and from the best fit parameters obtained by fitting a $\beta$ model to the X-ray surface brightness profile. The result obtained by integrating the density profile over a defined volume of the cluster depends on the cosmology assumed because the projected radius of the cluster is calculated using the cluster's angular distance $d_{A}$ (\citealt{Ferramacho:07}). 

The results of calculations are presented in Tab. \ref{tab1}. The main source of error in the determination of the gas mass comes from the normalization parameter in the spectral fit. This parameter is used to compute the gas density. In this way, the errors obtained in the gas mass come from fitting a single temperature model to the spectrum of a thermally inhomogeneous gas.

\section{Results}

Value $ L_{X} - T $ for distant clusters of galaxies, obtained for the cosmological parameters $ \Omega_{\Lambda} $ = 0.73, $ \Omega_{m} $ = 0.27, $ H_{0} $ = 73 km s$^{-1} $ Mpc$^{-1} $, is shown in Fig. \ref{lt}. Variation in the ratio is very small. It is almost entirely due to measurement errors. The slope is consistent with the ratio for clusters at low redshift. We used normalization parameter on $ z $ as $ L \sim (1 + z)^{A_{L_{X}T}} T^{\beta_{L_{X}T}} $ and we found that $ \beta_{L_{X}T} = $ 2.55 $\pm$ 0.07, and $ A_{L_{X}T} = 1.50 \pm 0.23 $ at confidence level 68 \%. 

Now we proceed to the consideration of dependencies, which concerns with the mass of hot intergalactic gas. Using spherically symmetric clusters, the value of gas mass is derived from surface brightness profile. It was decided to measure the mass of gas within radius $ R_{200} $.

The obtained dependence of $ M_{g,200} - T $ are given on Fig.~\ref{mt}. We used the normalization parameter as $ M_{g} \sim (1 + z)^{A_{M_{g}T}} T^{\beta_{M_{g}T}} $, as a result we found $ \beta_{M_{g}T} = $ 1.77 $\pm$ 0.16, and $ A_{M_{g}T} = -0.58 \pm 0.13 $. 

Relationship between mass and luminosity of gas is shown in Fig. \ref{lm}. This value is useful in assessing the functions of the distant clusters by their X-ray luminosity. The dependence of $ M_{g} - L_{X} $ has a strong evolutionary relationship at any values ​​of parameters. We used the normalization parameter as $ M_{g} \sim (1 + z)^{A_{M_{g}L_{X}}} L^ {\beta_{M_{g}L_{X}}} $, so we found $ \beta_{M_{g}L_{X}} = $ 0.73 $\pm$ 0.15, and obtained $ A_{M_{g}L_{X}} \approx -1.86 \pm 0.34$ for $ \Lambda $ = 0.73, $ \Omega_{m} $ = 0.27 Universe.

Thus the relationship between X-ray temperature, luminosity and mass of gas accumulations at $ z > $ 0.4 has a strong evolutionary relationship with respect to the correlations observed at low $ z $. The fact that, as already mentioned, the share of baryons in the mass accumulation of full $ f_{b} $ cause ought to be universal value, close to the average for the entire universe (\citealt{White:93}), but in this case the total mass of baryons can be found as $ M = M_{b}/f_{b} $. Accordingly, the evolution in the ratio $ M_{g} - L_{X} $ and $ M_{g} - T $ for baryon matter and non-baryon cause ought to be the same. 

The observed strong evolution based on $ M_{g} - L_{X} $ is weak, and negative evolution of the luminosity function means that the mass function evolves strongly: the spatial density of clusters of fixed mass at high $ z $ is much smaller than in the curent epoch of cosmological parameters (\citealt{Ettori:99}, \citealt{Ettori:00}). 

Supervisory evolutionary dependence in the ratio $ M_{g} - T - L_{X} $ indicates that the accumulation in the past had a great density - they were hotter and had a higher luminosity at fixed mass - that actually was predicted scaling theory of their formation (\citealt{Bryan:98}). 
\begin{figure}
 \centering
 \epsfig{file=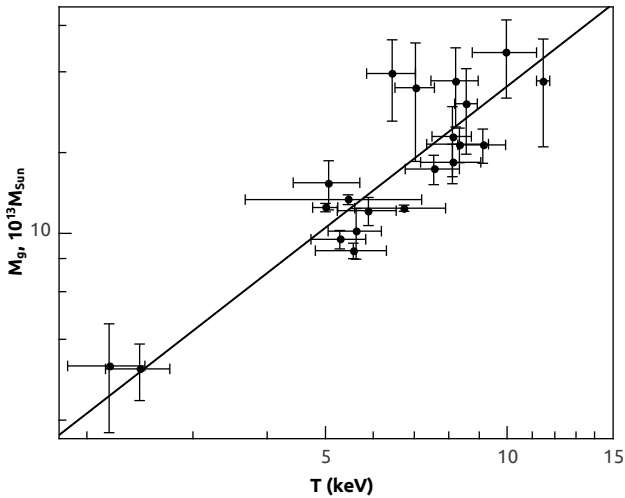,width=1.0\linewidth}
 \caption{The relation between temperature and mass of gas in radius, which correspond $\delta = $200. The solid line represents $M_{g} - T$ relation for near clusters.}\label{mt}
 \end{figure}

For our values ​​of cosmological parameters $\Omega$ and $\Lambda$ the $M_{g} - T$ relation varies considerably slower than expected from theory. Likely explanation for this discrepancy is the same as when considering the slope of the correlation $ L_{X} - T $ (\citealt{Frederiksen:09}). 

 \begin{figure*}
 \begin{minipage}{0.45\linewidth}
 \epsfig{file=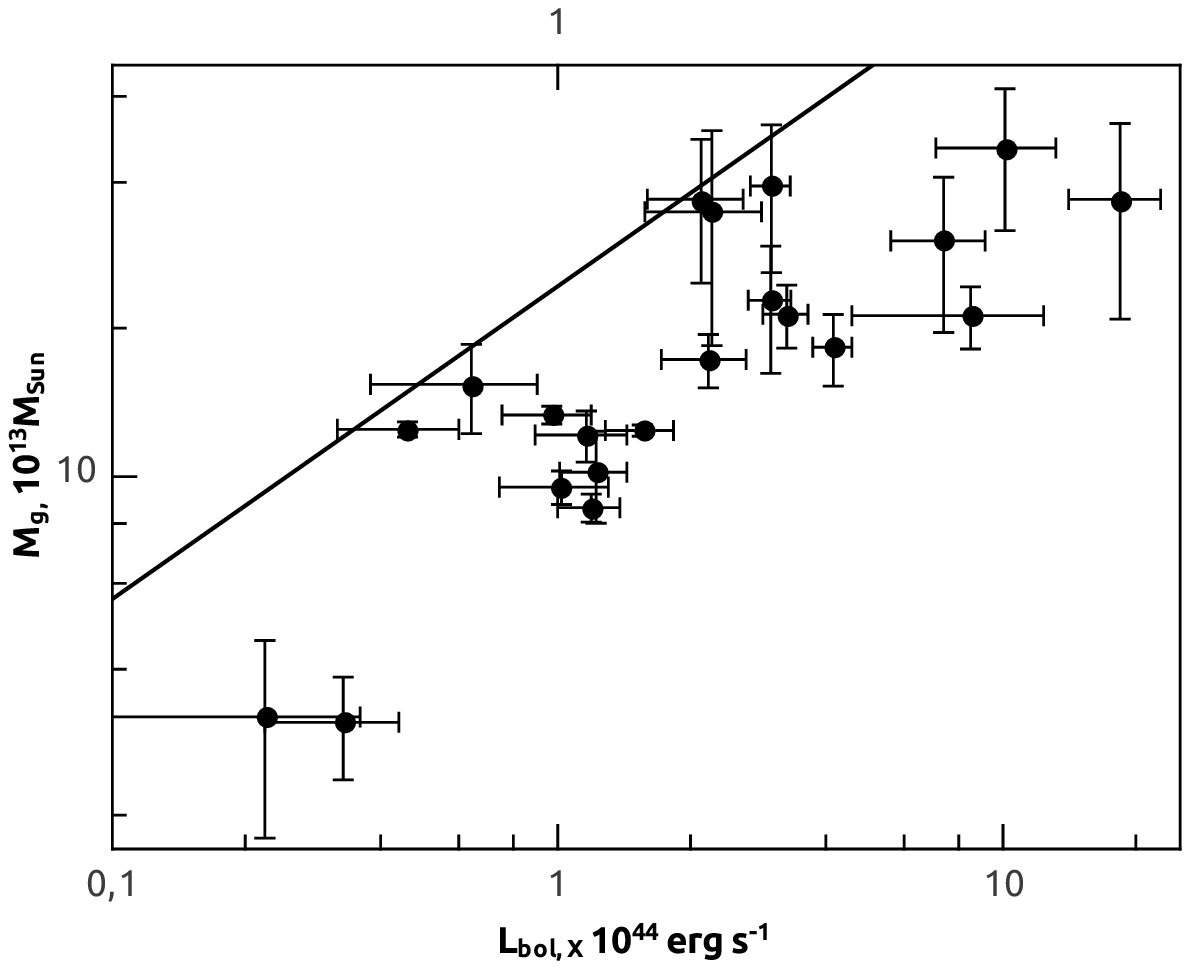, width=1.0\linewidth}
 \end{minipage}
 \begin{minipage}{0.45\linewidth}
 \epsfig{file=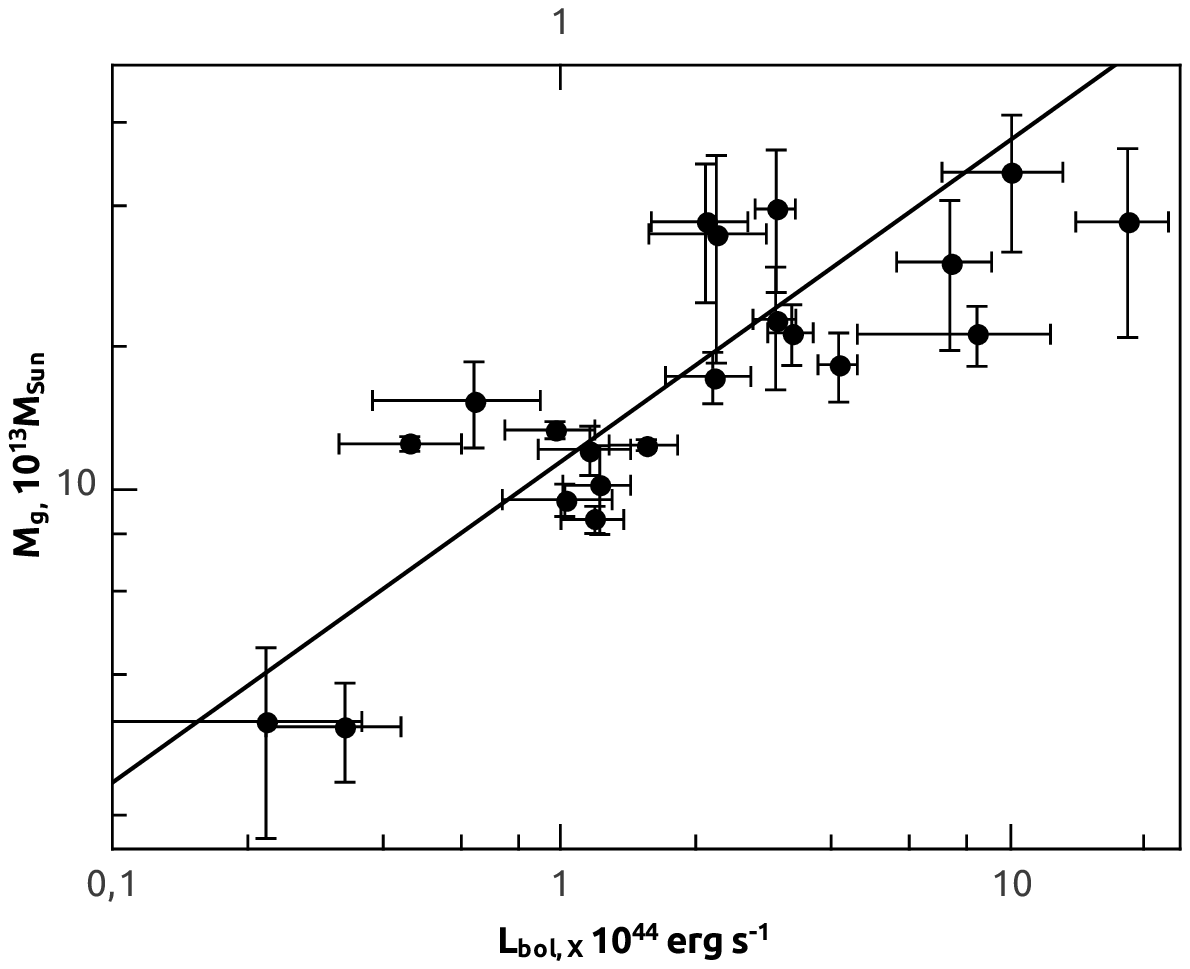, width=1.0\linewidth}
 \end{minipage}
 \caption{The relation between luminosity of clusters  in 0.5 - 7 keV energy range and mass of hot gas at $R_{200}$. The solid line on left image represents the $M - L_{X}$ relation for galaxy clusters on low redshift. The solid line on the right image moved on $(1+z)^{1.94}$. }\label{lm}
 \end{figure*}

\section{Discussion}

Using the new $Chandra$ cluster temperatures, luminosities and mass of X-ray hot intracluster gas, we obtain the $L_{X}-T-M_{X}$ relations in the 2 - 11 keV temperature interval with greatly reduced scatter. The large redshift sample has provided a evidence that clusters do not evolve self-similarly in the last $\sim$ 11 Gyr (approximately three quarter of the age of Universe). 

For all data we calculated the $\beta_{L_{X}T} = $ 2.55 $\pm$ 0.07. Using the X-ray luminosity function from the ROSAT Deep Cluster Survey in work (\citealt{Rosati:98}) and the Brightest Cluster Survey in paper (\citealt{Ebeling:97}), \citet{Borgani:99} found $ \beta_{L_{X}T}$ between 3 and 4. \citet{Mushotzky:97} found $A_{L_{X}T} = 0$ for all sample. In paper of \citet{Donahue:99}, authors found a slightly negative value of $A$ (for $\Omega_{m} = $0.3, $\Lambda = $0.7, $A_{L_{X}T} = -0.8_{-1.1}^{+0.9}$ and $\Omega_{m} = $ 1.0, $A_{LT} = -1.4_{-1.6}^{+0.8}$, $A_{L_{X}T} = 1.5$). \citet{Reichart:99} found the $L_{X} - T$ relation for function of cooling flow-corrected luminosities and temperatures. \citet{Reichart:99} like $\beta_{L_{X}T} = $2.80 $\pm $0.15 and $A_{L_{X}T} = 0.35_{-1.22}^{+0.54}$. These results in a good agreement with our results. \citet{Horner:99} determined for $\beta_{L_{X}T} = = 2.98 \pm 0.14$ and $A_{L_{X}T} = 0.02 \pm 0.16$ ($\Omega_{m} = 1.0$).

\citet{Sadat:98} found a positive evolution $A_{L_{X}T}$ = 0.5 $\pm$ 0.3 for $\Omega_{m}$ = 1.0, $\Lambda$ = 0.0 and \citet{Novicki:02} $A_{L_{X}T}$ = 1.1 $\pm$ 1.1 for $\Omega_{m}$ = 1.0, $\Lambda$ = 0.0, and $A_{L_{X}T}$ = 2.1 $\pm$ 1.0 for $\Lambda$ = 0.7, $\Omega_{m}$ = 0.3, respectively. \citet{Vikhlinin:02} found $A_{L_{X}T}$ = 0.6$\pm$0.3 for $\Lambda$ = 0.0, $\Omega_{m}$ = 1.0 and $A_{L_{X}T}$ = 1.5 $\pm$ 0.3 for a $\Lambda$ = 0.7, $\Omega_{m}$ = 0.3 cosmology.

Our normalization of the $M_{g} - T$ relation is also in a good agreement with those derived from recent high-resolution numerical simulations (e.g., \citealt{Borgani:04}) that attempted to model non-gravitational processes in the ICM (radiative cooling, star formation, and feedback from supernovas). The primary goal of observationally calibrating the $M_{g} - T$ relation is for use in fitting cosmological models to the cluster temperature function. We have presented the constraints on the mass-temperature relation of hot gas of galaxy clusters. 

\section*{Acknowledgments}
We thank the referee for a useful report which helped to improve the quality of the paper. This research has made use of data obtained from the Chandra Data Archive and the Chandra Source Catalog, and software provided by the Chandra X-ray Center (CXC) in the application packages CIAO, ChIPS, and Sherpa. We thank all the staff members involved in the Chandra project. The HEASARC online data archive at NASA/GSFC has been used extensively in this research. This research is particularly supported in frame of the ``Cosmo-Micro Physics'' program of the NAS of Ukraine.

\let\jnlstyle=\rm\def\jref#1{{\jnlstyle#1}}\def\aj{\jref{AJ}}
  \def\araa{\jref{ARA\&A}} \def\apj{\jref{ApJ}\ } \def\apjl{\jref{ApJ}\ }
  \def\apjs{\jref{ApJS}} \def\ao{\jref{Appl.~Opt.}} \def\apss{\jref{Ap\&SS}}
  \def\aap{\jref{A\&A}} \def\aapr{\jref{A\&A~Rev.}} \def\aaps{\jref{A\&AS}}
  \def\azh{\jref{AZh}} \def\baas{\jref{BAAS}} \def\jrasc{\jref{JRASC}}
  \def\memras{\jref{MmRAS}} \def\mnras{\jref{MNRAS}\ }
  \def\pra{\jref{Phys.~Rev.~A}\ } \def\prb{\jref{Phys.~Rev.~B}\ }
  \def\prc{\jref{Phys.~Rev.~C}\ } \def\prd{\jref{Phys.~Rev.~D}\ }
  \def\pre{\jref{Phys.~Rev.~E}} \def\prl{\jref{Phys.~Rev.~Lett.}}
  \def\pasp{\jref{PASP}} \def\pasj{\jref{PASJ}} \def\qjras{\jref{QJRAS}}
  \def\skytel{\jref{S\&T}} \def\solphys{\jref{Sol.~Phys.}}
  \def\sovast{\jref{Soviet~Ast.}} \def\ssr{\jref{Space~Sci.~Rev.}}
  \def\zap{\jref{ZAp}} \def\nat{\jref{Nature}\ } \def\iaucirc{\jref{IAU~Circ.}}
  \def\aplett{\jref{Astrophys.~Lett.}}
  \def\apspr{\jref{Astrophys.~Space~Phys.~Res.}}
  \def\bain{\jref{Bull.~Astron.~Inst.~Netherlands}}
  \def\fcp{\jref{Fund.~Cosmic~Phys.}} \def\gca{\jref{Geochim.~Cosmochim.~Acta}}
  \def\grl{\jref{Geophys.~Res.~Lett.}} \def\jcp{\jref{J.~Chem.~Phys.}}
  \def\jgr{\jref{J.~Geophys.~Res.}}
  \def\jqsrt{\jref{J.~Quant.~Spec.~Radiat.~Transf.}}
  \def\memsai{\jref{Mem.~Soc.~Astron.~Italiana}}
  \def\nphysa{\jref{Nucl.~Phys.~A}} \def\physrep{\jref{Phys.~Rep.}}
  \def\physscr{\jref{Phys.~Scr}} \def\planss{\jref{Planet.~Space~Sci.}}
  \def\procspie{\jref{Proc.~SPIE}} \let\astap=\aap \let\apjlett=\apjl
  \let\apjsupp=\apjs \let\applopt=\ao


\bsp
\label{lastpage}
\end{document}